\documentstyle[11pt,twoside,jltp]{article}
\def\trajectory#1#2#3#4{$T_{\rm i}=#1\,\rm K$, $P_{\rm i}=#2\,\rm bar$,
                         $T_{\rm f}=#3\,\rm K$, $P_{\rm f}=#4\,\rm bar$}

\title{Expansion of Liquid $^4$He Through the Lambda Transition}

\author{M.E. Dodd, P.C. Hendry, N.S. Lawson, P.V.E. McClintock
\address{Department of Physics, Lancaster University, LA1 4YB, UK}
and C.D.H. Williams\address{School of Physics, University of Exeter, EX4
4QL, UK}}

\runninghead{M.E. Dodd \it{\bf et al.}}
{Expansion of Liquid $^4$He Through the Lambda Transition}

\begin{document}

\begin{abstract} Zurek suggested ({\it Nature} {\bf 317}, 505; 1985)
that the Kibble mechanism, through which topological 
defects such as
cosmic strings are believed to have been created in the early Universe,
can also result in the formation of topological defects in liquid
$^{4}He$, i.e.\ quantised vortices, during rapid quenches through the
superfluid transition. Preliminary experiments (Hendry et al, Nature
{\bf \itshape 368}, 315; 1994) seemed to support this idea in that the
quenches produced the predicted high vortex-densities. The present
paper describes a new experiment incorporating a redesigned expansion
cell that minimises vortex creation arising from conventional
hydrodynamic flow. The post-quench line-densities of vorticity produced
by the new cell are no more than $10^{10}m^{-2}$, a value that is at
least two orders-of-magnitude less than the theoretical prediction. We
conclude that most of the vortices detected in the original experiment
must have been created through conventional flow processes.

PACS numbers: 11.27.+d, 05.70.Fh, 11.10.Wx, 67.40.Vs
\end{abstract}

\maketitle


\section{INTRODUCTION} After a physical system has passed rapidly
through a continuous phase-transition, its order-parameter can have
components with large differences between adjacent, but causally
disconnected, ``domains''. In such systems topological defects can form
at the domain boundaries\cite{A}. This idea was proposed by Kibble
\cite{B} in connection with the grand unified theory (GUT)
symmetry-breaking phase-transition of the early Universe, and has been
developed by Zurek \cite{C,D,E} who has estimated how the density of
defects created depends on the rate at which the system passes through
the transition. Zurek also pointed out that this mechanism of defect
production was applicable, in principle, to all continuous
phase-transitions, and that it should therefore be possible to validate
some aspects of cosmological theories through laboratory-scale
experiments. The first of these examined weakly first-order
phase-transitions in liquid crystals \cite{F,G}. Later, the
corresponding experiments were carried out using the second-order
superfluid phase-transitions of liquid $^4$He\ \cite{H} and liquid
$^3$He \cite{I,J}. All these experiments produced defect densities
reported as being consistent with Zurek's estimates\cite{C,D,E}.

In this paper we describe an improved version of the $^4$He experiment
\cite{H} in which particular care has been taken to minimise the
production of topological defects ({\it i.e.} quantised vortices) by
ordinary hydrodynamic flow processes. Our new results, of which a
preliminary report\cite{newprl} has already been published, show no
convincing evidence of {\it any} vortex creation at all. But they allow
us to place an approximate upper-bound on the initial density of
vortices produced by the Kibble-Zurek mechanism.

\section{THEORETICAL BACKGROUND} A considerable amount is known about
the properties\cite{S,T,O,K} of vortices, and about how they are
created at very low temperatures\cite{avenel,muirhead,mqt} but the
mechanism responsible for the vorticity that appears\cite{W,H} as a
result of passing through the $\lambda$-transition (which separates the
normal \hbox{helium-I} and the superfluid \hbox{helium-II} phases) is
not understood. One possibility is that pre-existing rotational flow
caused by e.g.\ convection or boiling in the \hbox{helium-I} phase is,
with the onset of long-range order, converted into quantised vortex
lines in \hbox{helium-II}; this is quite distinct from the Zurek
scenario which we shall now briefly describe.

\begin{figure}[h]
\vspace{6in}
\includegraphics{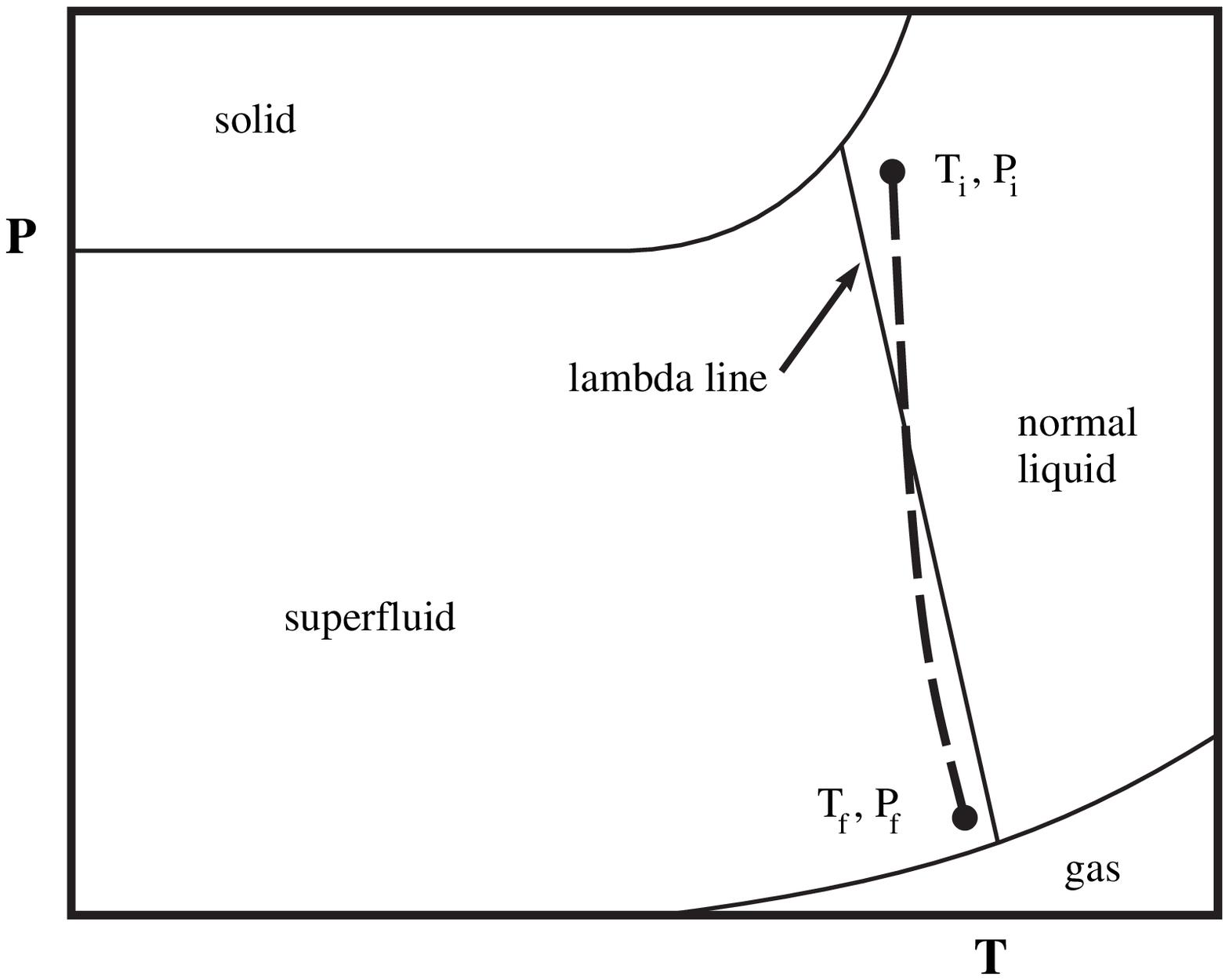}
\caption{Schematic of expansion trajectory through the $^4$He superfluid
transition from a starting temperature and pressure $(T_i, P_i)$ to final
values $(T_f, P_f$).}
\label{trajectories} \end{figure}

The underlying idea \cite{C} is quite simple. A small isolated volume
of \hbox{helium-I} is initially held at pressure $P_{\rm i}$, and
temperature $T_{\rm i}$, just above the temperature $T_\lambda(P_{\rm
i})$ of the $\lambda$-transition. The logarithmic infinity in its heat
capacity at $T_\lambda$ makes it impossible to {\it cool} the sample
quickly into the superfluid phase but the pressure dependence of
$T_\lambda$ means that it can be taken through the transition very
rapidly by adiabatic {\it expansion} into the \hbox{helium-II} phase,
to a final pressure $P_{\rm f}$ and temperature $T_{\rm f}$
(Fig~\ref{trajectories}). Fluctuations present in the
\hbox{helium-I} are expected to cause the nascent superfluid to form
with a spatially incoherent order-parameter, corresponding to a large
density of vortex lines. This scenario depends on the fact that the
liquid can, in principle, expand at velocities comparable to that of
first-sound, whereas the propagation velocity for changes in the
order-parameter is limited by the much slower velocity of second-sound.

\begin{figure}[h]
\vspace{7in}
\includegraphics{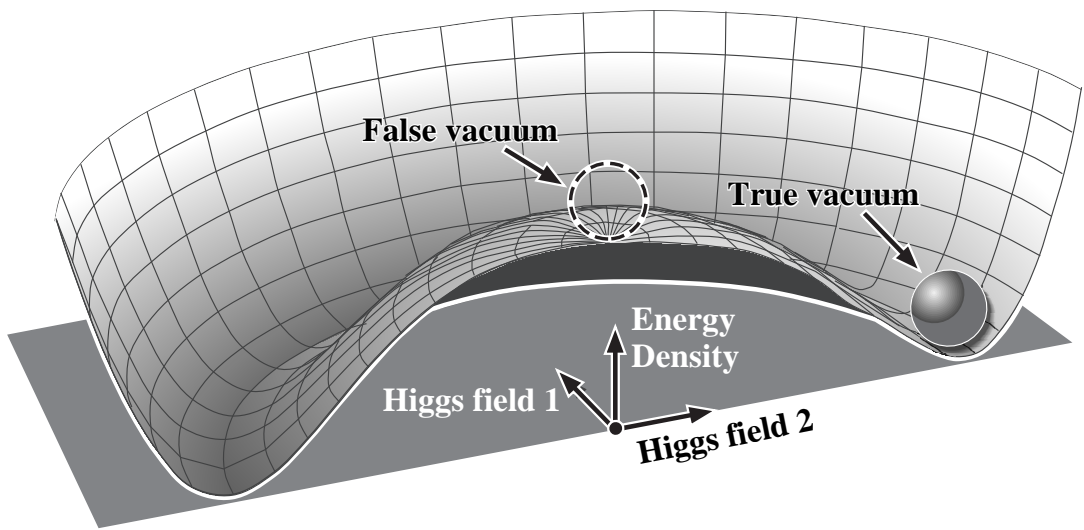}
\caption{Potential contribution to the free energy for the cosmological
phase transition, after Guth and Steinhardt$^{22}$.
}
\label{sombrero} \end{figure}

The analogy \cite{C,D,E} between liquid helium and the early Universe
arises because they both be considered to undergo second-order phase
transitions describable in terms of Ginsburg-Landau theory \cite{X}. In
each case the potential contribution to the free-energy density can be
written as:

\begin{equation}
V = \alpha(T)|\psi|^2+{1 \over 2}\beta|\psi|^4
\end{equation}

\noindent where the parameter $\alpha$ is positive at temperatures
above $T_\lambda$ and negative below it, and $\beta$ is a
constant. For liquid $^4$He the order-parameter is the
modulus of the complex-scalar field $\psi$, {\it i.e.} the
Bose condensate wave-function which is a solution of the
Ginsburg-Pitaevskii equation \cite{X}. In the cosmological
analogy the components of $\psi$ are Higgs fields
\cite{D,E}. In the symmetric (\hbox{helium-I}) phase
$T>T_\lambda$ and the time-average of the order-parameter
$\langle\psi\rangle=0$. Below $T_\lambda$ this gauge
symmetry, is broken so $\langle\psi\rangle$ becomes
non-zero, and the real and imaginary parts of the potential
in equation~1 acquire the same ``sombrero'' shape as the
corresponding cosmological free energy expressed in terms of
Higgs fields (Fig.~\ref{sombrero}). In the early Universe
a symmetry-breaking phase-transition from a false-vacuum
state to a true-vacuum state is thought to have occurred
once the temperature had fallen to $\sim10^{27}\,\rm K$,
about $10^{-35}\,\rm s$ after the big bang. Although there
are many variants of the basic model, with and without
inflation, it is believed that a variety of topological
defects \cite{B} would have been produced in the transition
because an event horizon prevented adjacent regions from
being causally connected. Cosmic strings \cite{L} --- thin
tubes of false vacuum --- are one such defect and may have
had a role in galaxy formation. It is these that
correspond to the quantised vortices found in
\hbox{helium-II}.  The analogy between the helium and cosmological
phase transitions may thus be summarised as follows --

\begin{center}
\vspace{0.2cm}

\begin{tabular}{lcl}
Higgs field 1 & $\longleftrightarrow$ & Re $\psi$\\
Higgs field 2 & $\longleftrightarrow$ & Im $\psi$\\
False vacuum  & $\longleftrightarrow$ & He I\\
True vacuum   & $\longleftrightarrow$ & He II\\
Cosmic string & $\longleftrightarrow$ & quantized vortex line
\end{tabular}

\end{center}
\vspace{0.2cm}

\section{DETECTING VORTICITY}
In our initial experiments \cite{H,M}, the vortex density was
measured by recording the attenuation of a sequence of
second-sound pulses propagated through the \hbox{helium-II}.
We expected that, following an expansion, the pulse
amplitude would grow towards its vortex-free value as the
tangle decayed and the attenuation decreased.

A considerable amount is known about the decay \cite{T,O,K,U} of
hydrodynamically created vortex tangles in \hbox{helium-II}. Numerical
simulations \cite{U,V} give a good qualitative description of the
manner in which a homogeneous isotropic tangle evolves and decays. The
rate at which it occurs in this temperature range is governed by the
Vinen \cite{O} equation

\begin{equation}
\frac{{\rm d}L}{{\rm d}t} = - \chi_2 \frac{\hbar}{m_4} L^2
\label{eqnVinen} \end{equation}

\noindent where $L$ is the vortex-line density at time $t$, the $^4$He
atomic mass is $m_4$, and $\chi_2$ is a dimensionless parameter. The
relationship between vortex density and second-sound attenuation is
known \cite{K} from experiments with rotating helium, and may for
present purposes be written

\begin{equation} L = \frac{6 c_2}{\kappa B d}\ln (S_0/S) \label{eqnL}
\end{equation} 

\noindent where $c_2$ is the velocity of second-sound, $S$ and $S_0$
are the signal amplitudes with and without vortices present
respectively. $B$ is a weakly temperature dependent parameter, $\kappa
= h/m_4$ is the quantum of circulation, and $d$ is the transducer
separation. Integrating equation~\ref{eqnVinen} and substituting for
$L$ from equation~\ref{eqnL} gives an expression for the recovery of
the signal: 

\begin{equation} {1 \over {\ln(S_0)-\ln(S)}} =
\frac{6c_2}{\kappa B d} \left( \frac{\chi_2\kappa t}{2\pi} +
\frac{1}{L_1} \right) \label{eqnLine} \end{equation}

\noindent where $L_1$ is the vortex density immediately after the
expansion. All the constants in this expression are known, although
$\chi_2$ and $B$ do not seem to have been measured accurately within
the temperature range of interest.

\section{THE FIRST EXPERIMENT}
 
\begin{figure}[h]
\vspace{6in}
\includegraphics{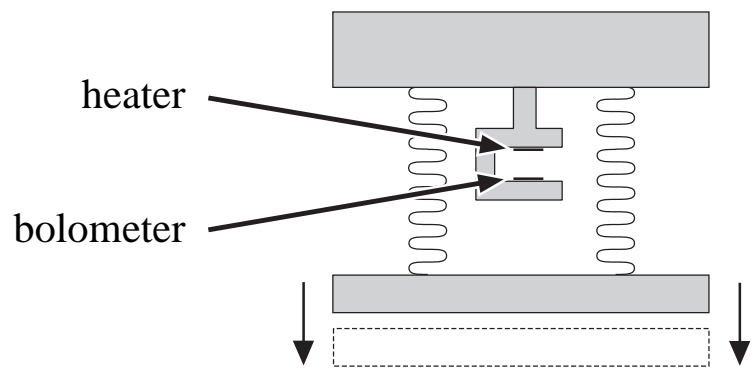}
\caption{Schematic diagram showing the main features of the original
experimental cell. The heater and bolometer were mounted on a yoke
immersed in the liquid $^4$He.}
\label{original} \end{figure}

A description of our first attempt to realise a bulk version \cite{D}
of Zurek's experiment has been given in a previous paper \cite{M} but,
briefly, the arrangement was as follows. A cell with phosphor-bronze
concertina walls was filled, by condensing in isotopically\cite{Z} pure
$^4$He through a capillary tube, and then sealed with a
needle-valve (Fig.~\ref{original}). The top of the cell was fixed
rigidly to the cryostat but its bottom surface could be moved to
compress the liquid, or released to expand it, using a pull-rod from
the top of the cryostat. The cell was {\it in vacuo} surrounded by a
reservoir of liquid $^4$He at $\sim 2\,\rm K$. A Straty-Adams
capacitance gauge \cite{ee} recorded the pressure in the cell and
carbon-resistors were used as thermometers, one in the reservoir, one
on the cell. The temperature of the cell could be adjusted by means of
an electrical heater and a breakable thermal link to the $2\,\rm K$
reservoir. A trigger mechanism on the mechanical linkage allowed the
cell to increase its volume by $\sim 20\%$ very rapidly under the
influence of its own internal pressure.

\begin{figure}[h]
\vspace{7.8in}
\includegraphics{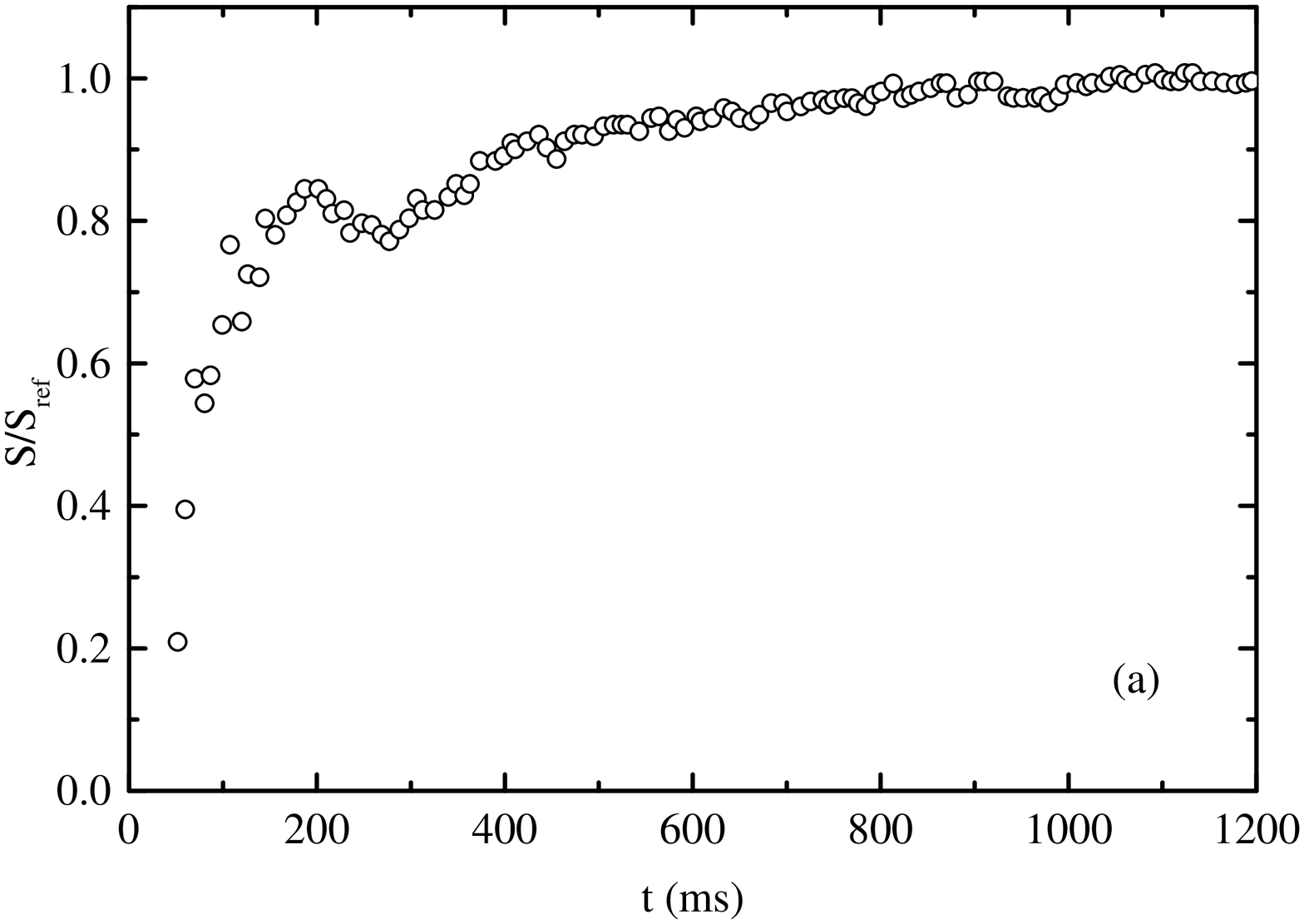}
\includegraphics{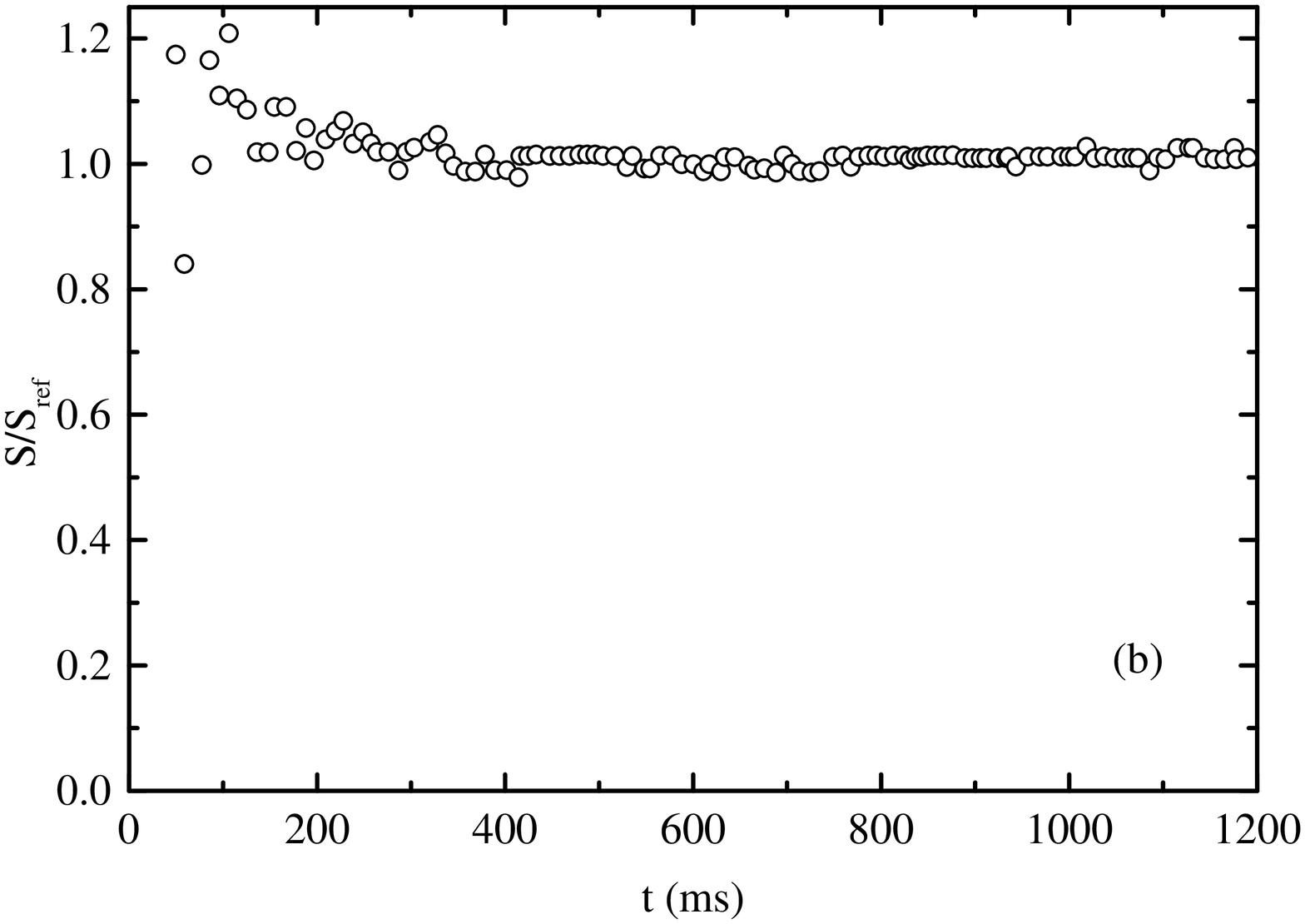}
\includegraphics{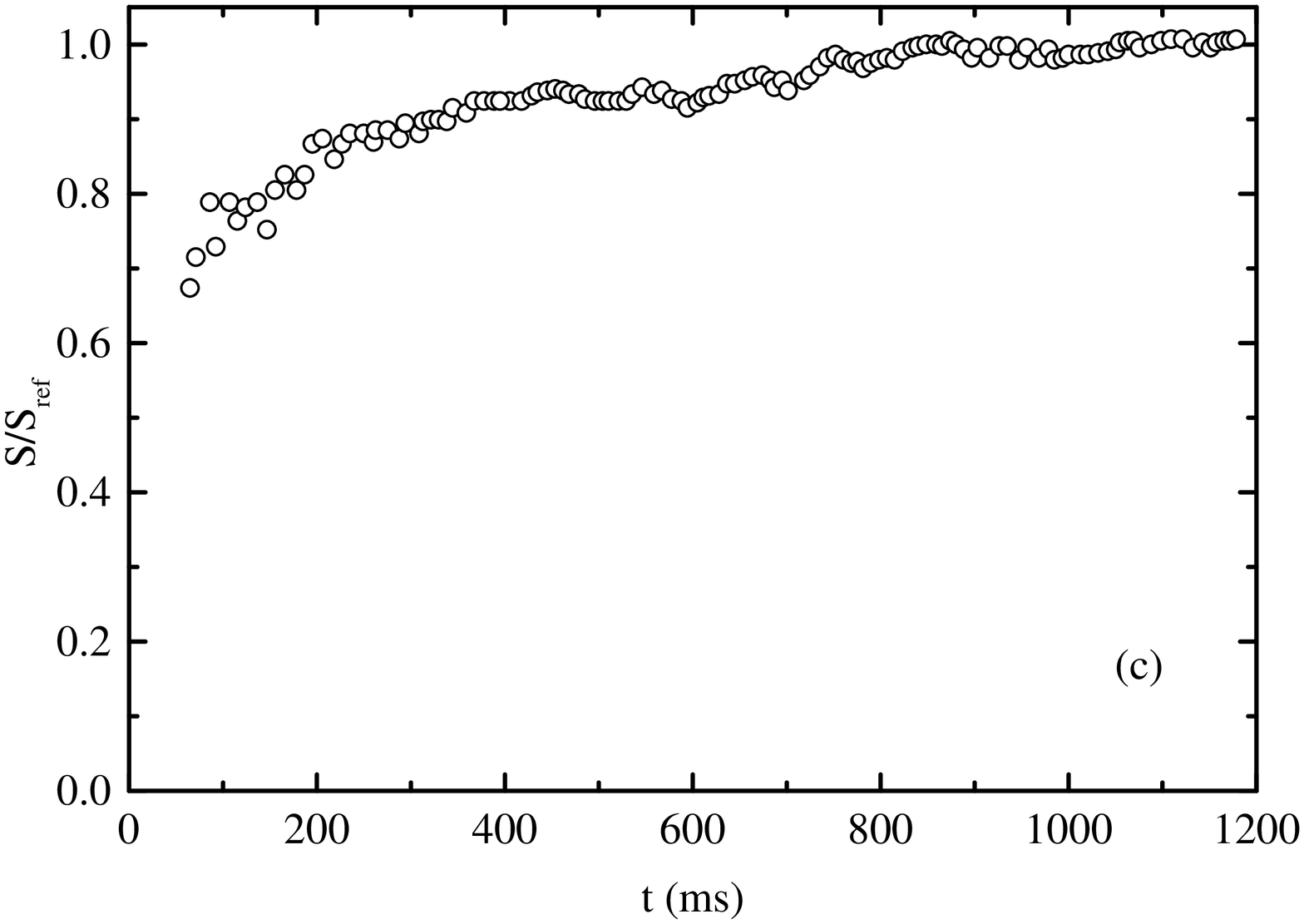}
\caption{Recovery of the scaled second-sound signal amplitude $S$
as a function of time $t$ in our original experiment, for various
expansion trajectories: (a) through the lambda line,
\trajectory{1.81}{29.6}{2.05}{6.9};
(b) starting well below the lambda line,
\trajectory{1.58}{23.0}{1.74}{4.0};
(c) starting slightly ($\sim 10\,\rm mK$)  below the lambda line,
\trajectory{1.82}{25.7}{2.03}{6.9}. }
\label{recovery} \end{figure}

As described above, the vortex density was inferred from the amplitude
of second-sound pulses created with a thin-film heater. The signal was
detected with a bolometer and passed, {\it via} a cryogenic FET
preamplifier, to a Nicolet 1280 data processor which recorded some
$\sim200$ pulses during 1--2 s following the expansion. The last pulses
in the sequence define $S_0$, the signal amplitude in the (virtual)
absence of vortices \cite{bb}. Fig.~\ref{recovery} shows examples of
data recorded with this first version of the experiment.

There was a ``dead period'' of about $50\,\rm ms$ after the mechanical
shock of the expansion which caused vibrations that obscured the
signals. The initial vortex-density was therefore obtained from plots
such as Fig.\ \ref{recovery}(a) by back-extrapolation to the moment $t=0$ of
traversing the transition and was found to be $\sim
10^{12}-10^{13}\,{\rm m}^{-2}$ consistent with the theoretical
expectation \cite{C,D,E}. However, an unexpected observation in these
initial experiments \cite{H} was that small densities of vortices were
created even for expansions that occurred wholly in the superfluid
phase (Fig.~\ref{recovery}c), provided that the starting point was
very close to $T_\lambda$. The phenomenon was initially \cite{M}
attributed to vortices produced in thermal fluctuations within the
critical regime, but it was pointed out \cite{N} that effects of this
kind are only to be expected for expansions starting within a few
microkelvin of the transition, {\it i.e.} much closer than the typical
experimental value of a few millikelvin. The most plausible
interpretation --- that the vortices in question were of conventional
hydrodynamic origin, arising from non-idealities in the design of the
expansion chamber --- was disturbing, because expansions starting above
$T_\lambda$ traverse the same region. Thus some, at least, of the
vortices seen in expansions through the transition were probably not
attributable to the Zurek-Kibble mechanism as had been assumed. It has
been of particular importance, therefore, to undertake a new experiment
with as many as possible of the non-idealities in the original design
eliminated or minimised.

\section{THE NEW EXPERIMENT}
An ideal experiment would avoid all fluid flow parallel to surfaces
during the expansion. This could, in principle, be accomplished by the
radial expansion of a spherical volume, or the axial expansion of a
cylinder with stretchy walls. In neither of these cases would there be
any relative motion of fluid and walls in the direction parallel to the
walls, and hence no hydrodynamic production of vortices. However, it is
impossible to eliminate the effects of fluid-flow completely from any
real apparatus. At the $\lambda$-transition the critical velocity tends
to zero, so {\it any} finite flow-velocity will create vorticity.
However, the period during which this happens is very short; the entire
expansion takes only a few milliseconds so only limited growth can
occur from the surface-vorticity sheet of half-vortex-rings \cite{cc}.

With hindsight, we can identify the principal causes of unintended
vortex creation in the original experiment\cite{H,M}, in order of
importance, as follows: (a) expansion of liquid into the cell from the
filling capillary, which was closed by a needle valve $0.5\,\rm m$
up-stream; (b) expansion of liquid out of a shorter capillary
connecting the cell to the Straty-Adams capacitive pressure-gauge; (c)
flow out of and past the fixed yoke (a U-shaped structure) on which the
second-sound transducers were mounted. In addition, (d) there were the
undesirable transients caused by the expansion system bouncing against
the mechanical stop at room temperature. The walls of the cell were
made from bronze bellows \cite{H,M}, rather than being a stretchy
cylinder, but the effects of the flow parallel to the convoluted
surfaces were relatively small.

\begin{figure}[h]
\vspace{6in}
\includegraphics{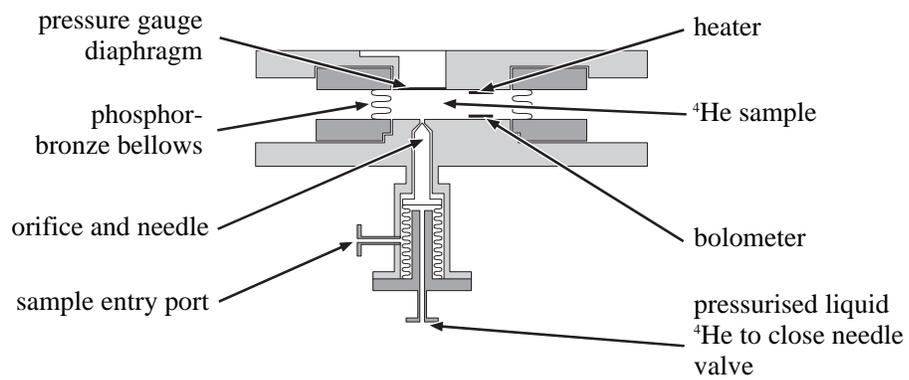}
\caption{The new expansion cell, designed so as to minimise hydrodynamic
creation of vortices.}
\label{redesigned}
\end{figure}

The design of our new expansion cell (Fig.~\ref{redesigned})
addressed all the listed problems, as follows: (a) a
hydraulically-operated needle-valve eliminated the dead-volume of
capillary tube; (b) the phosphor-bronze diaphragm of the pressure gauge
\cite{ee} became an integral part of the upper cell-wall eliminating
the long tube leading to the pressure gauge; (c) the cell was shortened
from $25 \rm mm$ to $\sim 5 \rm mm$ so that the heater-bolometer pair
could be mounted flush with its top and bottom inner surfaces, thereby
eliminating any need for a yoke; (d) some damping of the expansion was
provided by the addition of a (light motor-vehicle) hydraulic
shock-absorber.

The operation of the apparatus, the technique of data collection and
the analysis were much as described previously \cite{H,M} except that the
rate at which the sample passed through the $\lambda$-transition was
determined directly by simultaneous measurements of the position
\cite{aa} of the pull-rod (giving the volume of the cell, and hence its
pressure) and the temperature of the cell.

The position detector made use of the variation of inductance of a
solenoid coil when a ferrite-core is moved in and out. The 1 cm long, 4
mm diameter, ferrite-core was rigidly attached to the top of the
pull-rod, where it emerged from the cryostat, and was positioned so
that half of it was inside the 5 mm bore of the solenoid. The solenoid
itself was clamped to the cryostat top-plate so that, as the cell
expanded and the pull-rod moved down, the ferrite penetrated further
inside. The position was measured by making the solenoid part of a
resonant LCR circuit and setting the frequency so that the output was
at half maximum. The 2 MHz AC signal was rectified, smoothed and then
measured using a digital oscilloscope. The response time, which was
limited by the smoothing and the $Q$ of the circuit, was $\sim 50 \mu$s
and the sensitivity was $\sim$500 mV/mm. This, and the linearity of the
system, were measured by calibration against a micrometer. The position
could in principle be measured to an accuracy of $\sim \pm 2 \mu$m but
in the experiment this became $\sim \pm 10 \mu$m because of the 8 bit
digitiser of the oscilloscope.

We define the
dimensionless distance from the transition by
\begin{equation}
\epsilon = \frac{T_{\lambda} - T}{T_{\lambda}}
\label{epsilon}
\end{equation}
and the quench time $\tau_{\rm q}$ by
\begin{equation}
\frac{1}{\tau_{\rm q}} = {\left(\frac{{\rm d}\epsilon}{{\rm d}t}
\right)_{\epsilon = 0}}
\label{quench}
\end{equation}

\begin{figure}[h]
\vspace{8in}
\includegraphics{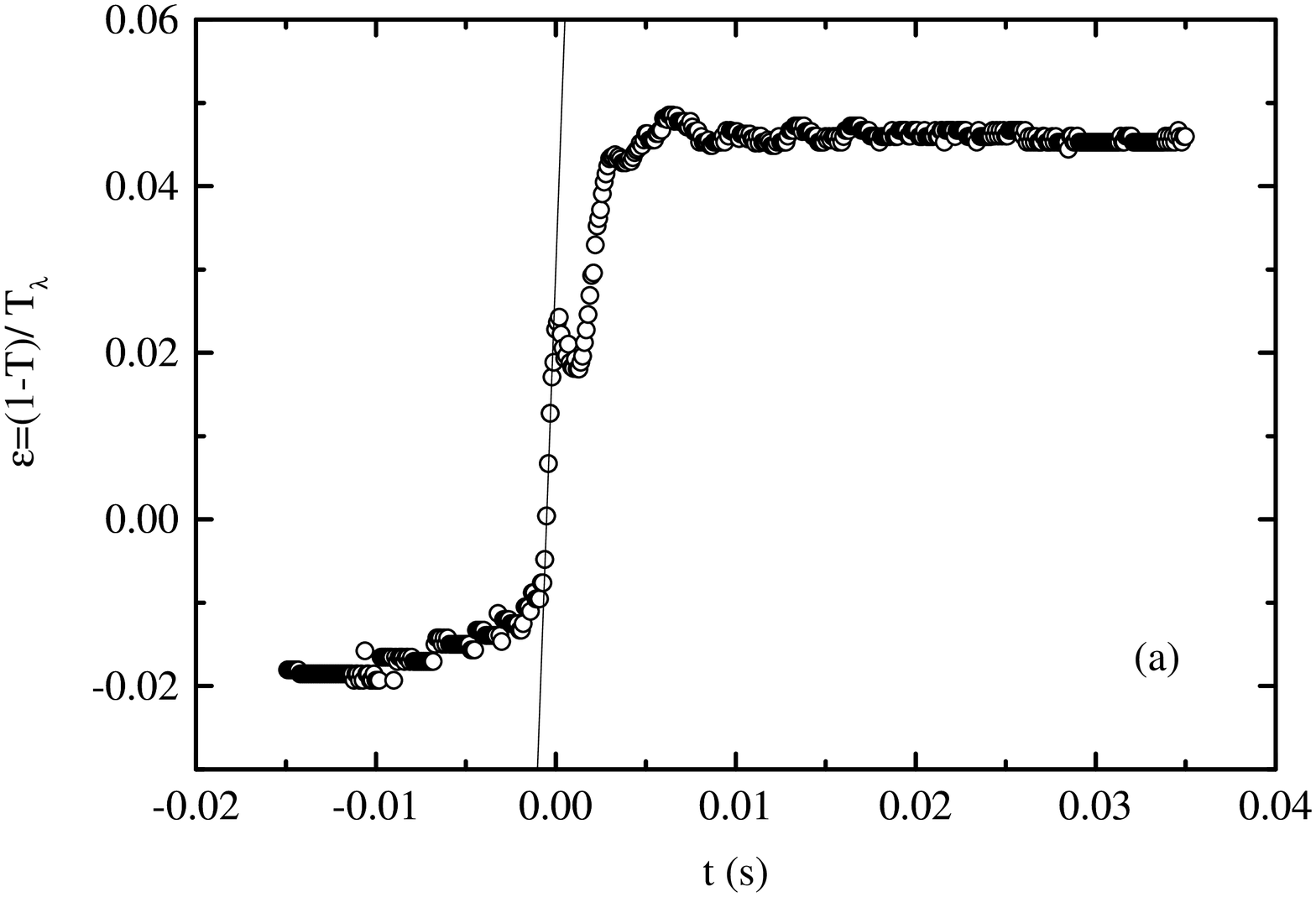}
\includegraphics{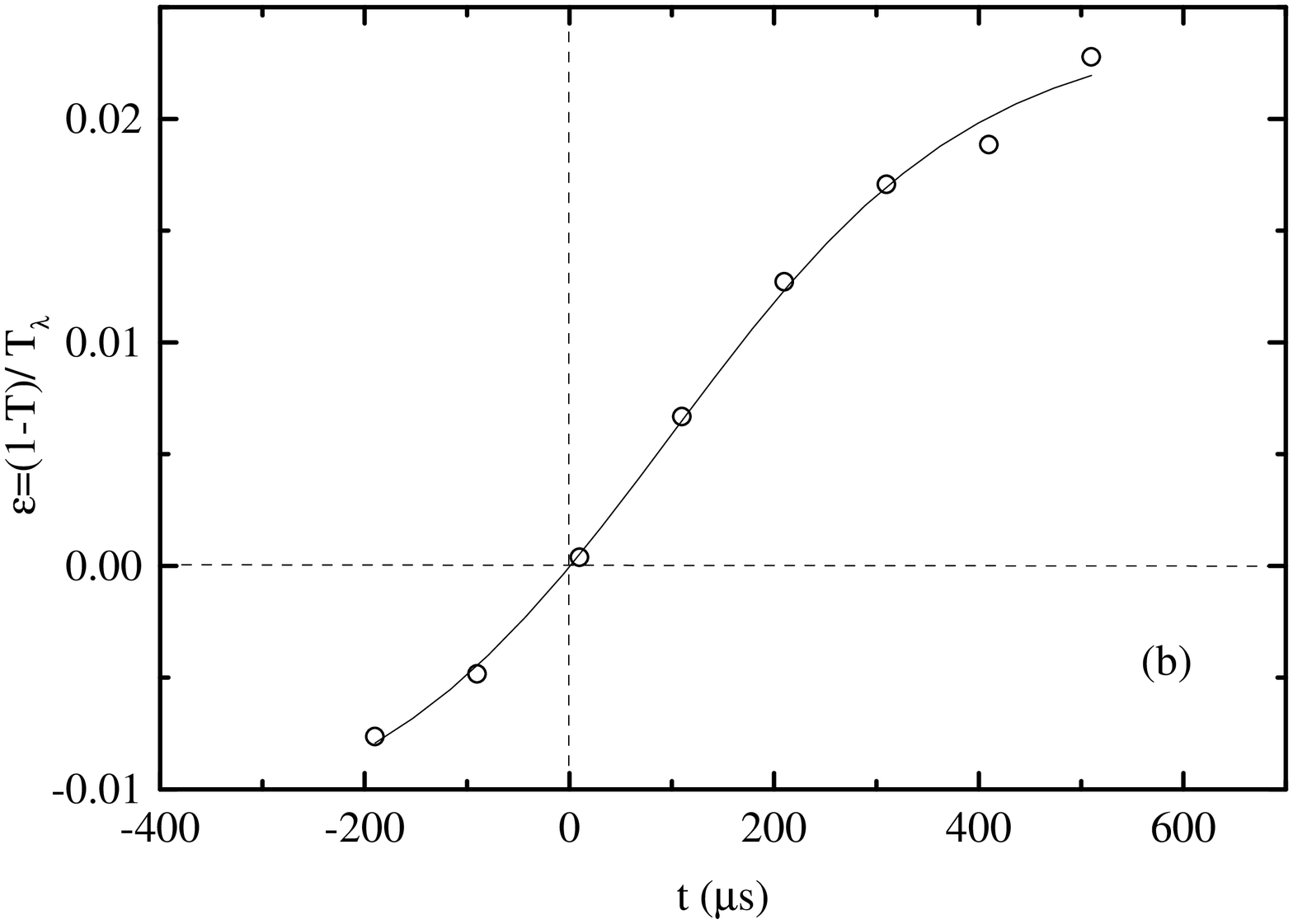}
\caption{Distance {\it versus} time $t$ for a typical quench: (a) during
the complete expansion; (b) enlargement of region near $\epsilon = 0$.
 The reciprocal of the gradient at the transition gives the value of
the `quench time' parameter, $\tau_{\rm q}$.}
\label{Fig6}
\end{figure}

\noindent Fig.~\ref{Fig6} shows a typical evolution of $\epsilon$ with time
during an expansion. In Fig.~\ref{Fig6}(a), which plots the full
expansion period, it is evident that the system ``bounces'' momentarily
near $\epsilon = 0.02$, but without passing back through the
transition. The effect is believed to be associated with the onset of
damping from the shock absorber, after backlash in the system has been
taken up.

The quench time is readily determined from $\epsilon (t)$ near the
transition. In the case illustrated in Fig.~\ref{Fig6}, shown in
expanded form in part (b), it was $\tau_{\rm q}=17\pm 1\,\rm ms$. We
are thus able take both the pressure-dependence of $T_\lambda$, and the
non-constant rate of expansion, explicitly into account.

\section{RESULTS FROM THE NEW CELL}
\begin{figure}[h]
\vspace{7.8in}
\includegraphics{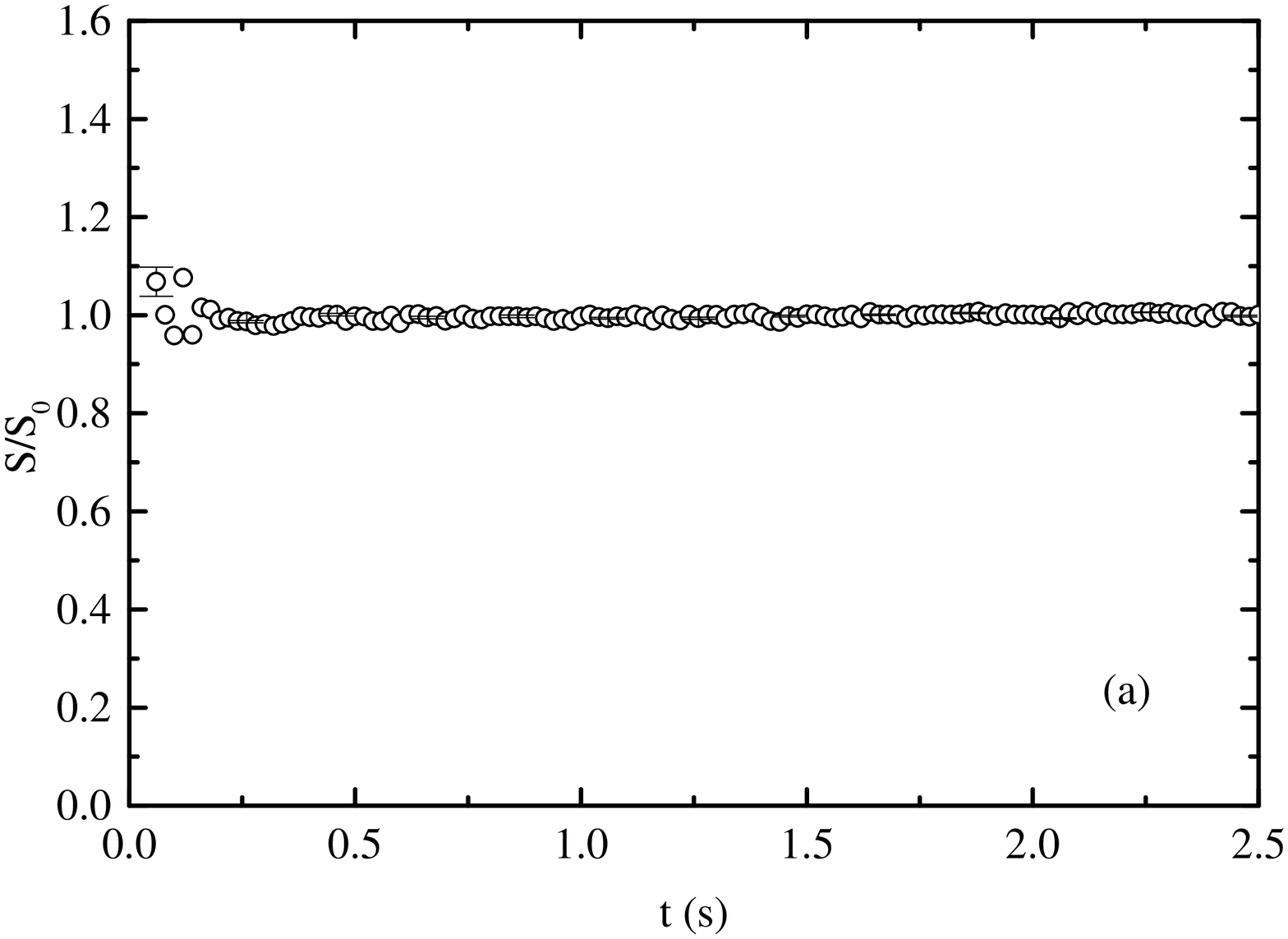}
\includegraphics{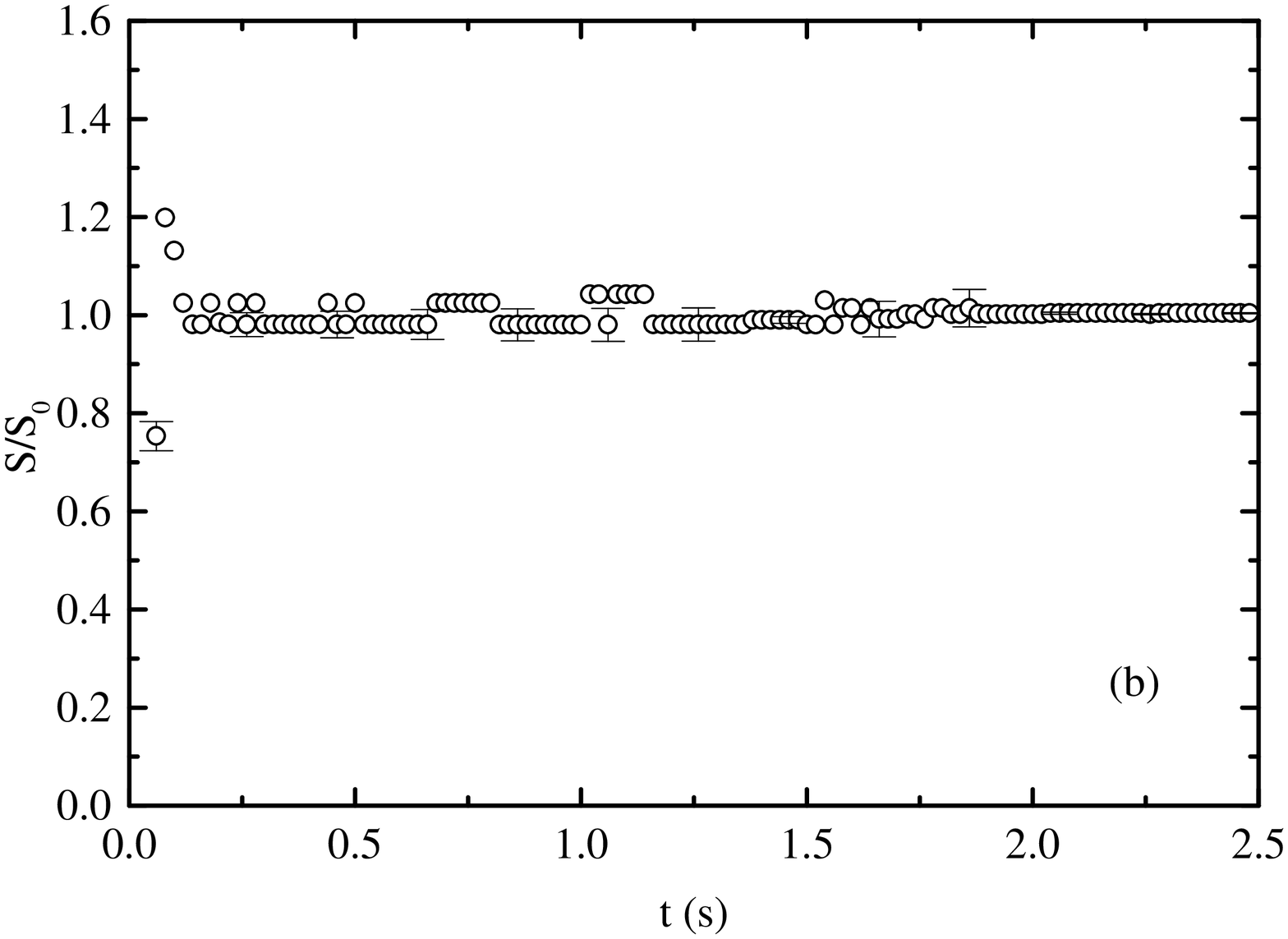}
\includegraphics{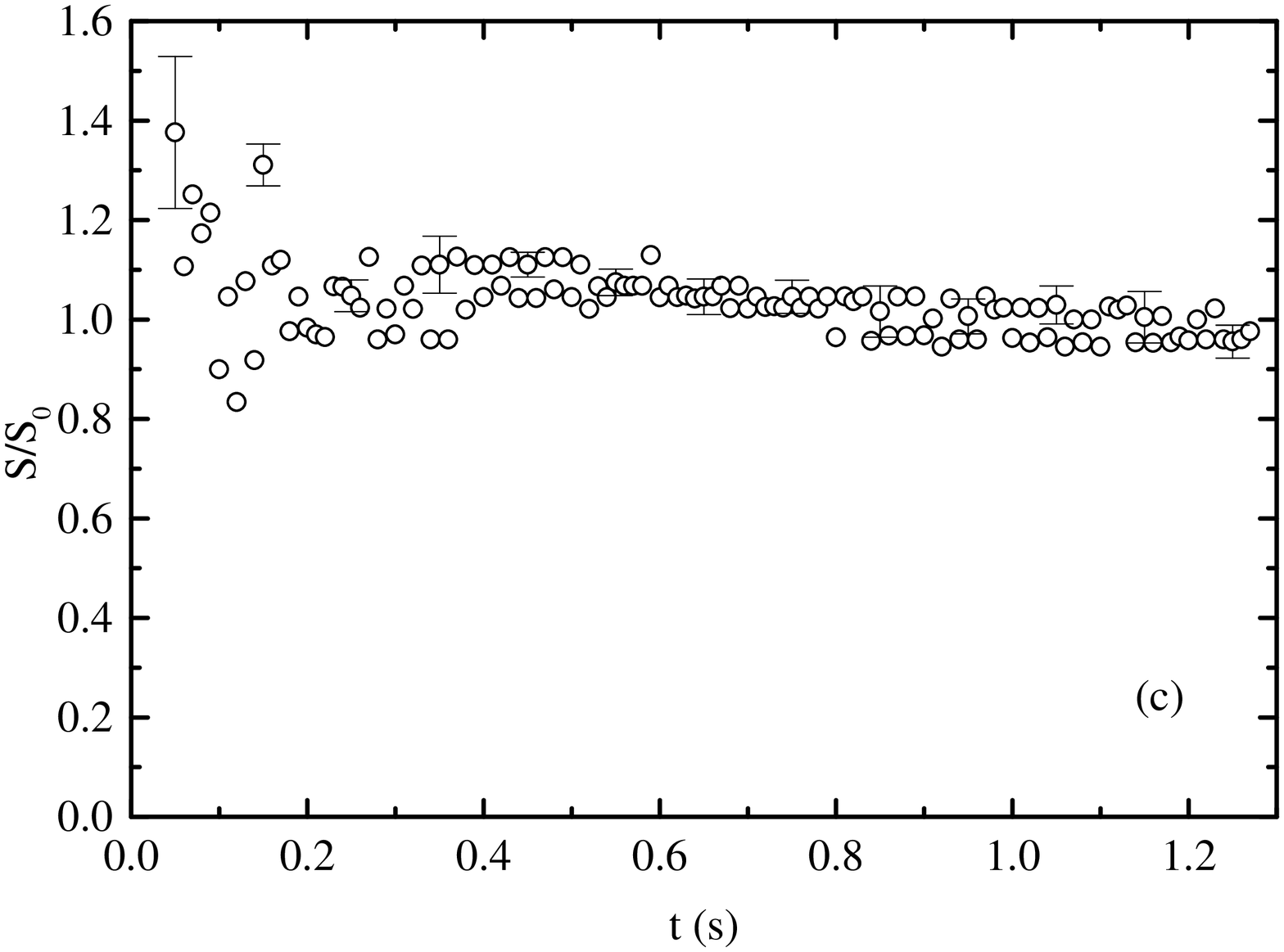}
\caption{Second-sound pulse amplitude $S/S_0$ as a function of time
following pressure quenches entirely within the helium-II phase:
(a) starting far ($\sim 490\,\rm mK$) below $T_\lambda$ with
\trajectory{1.37}{24.2}{1.47}{7.2};
(b) starting $\sim 150\,\rm mK$ below $T_\lambda$ with
\trajectory{1.74}{23.8}{1.76}{6.9};
(c) starting slightly ($\sim 40\,\rm mK$) below $T_\lambda$ with
\trajectory{1.85}{22.4}{1.98}{6.4}. }
\label{Fig7}
\end{figure}

The improvements in the cell-design intended to eliminate vortices
originating from conventional fluid-flow were clearly successful; as we
had hoped, and unlike the first cell, expansion trajectories that stay
within the superfluid phase, even those starting as close as $42\,\rm
mK$ to the transition, create no detectable vorticity
(Fig.~\ref{Fig7}).

\begin{figure}[h]
\vspace{7.8in}
\includegraphics{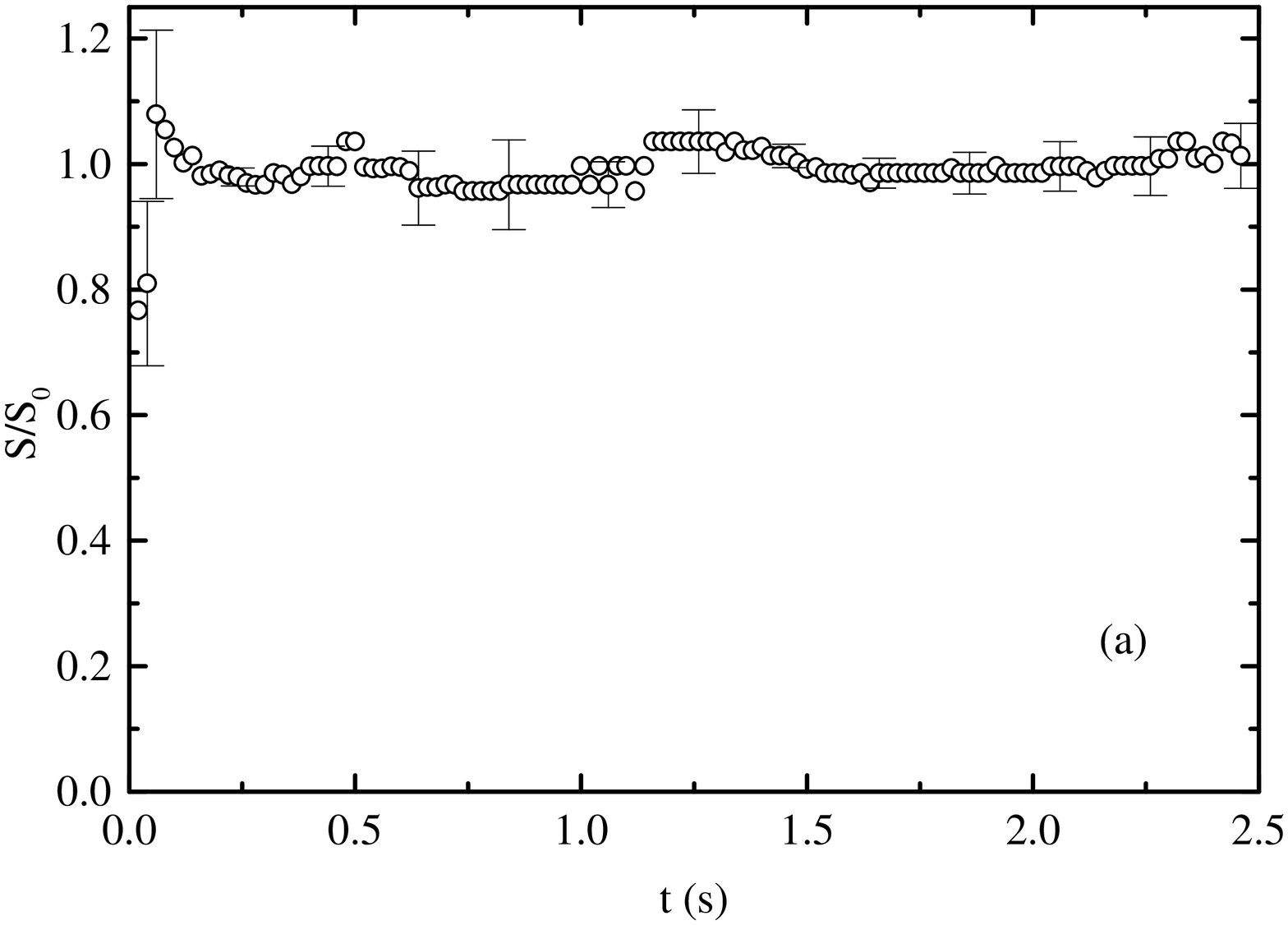}
\includegraphics{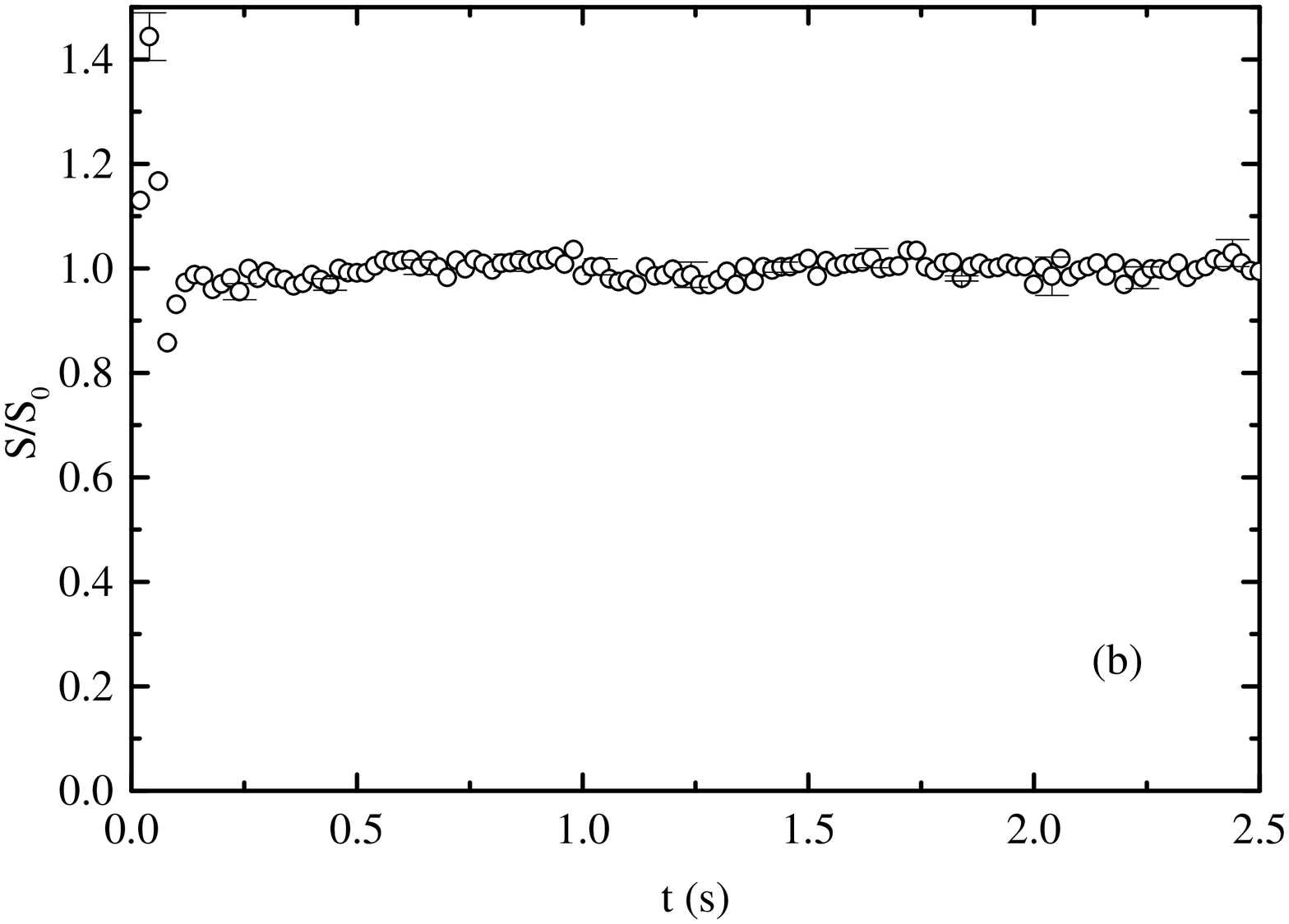}
\includegraphics{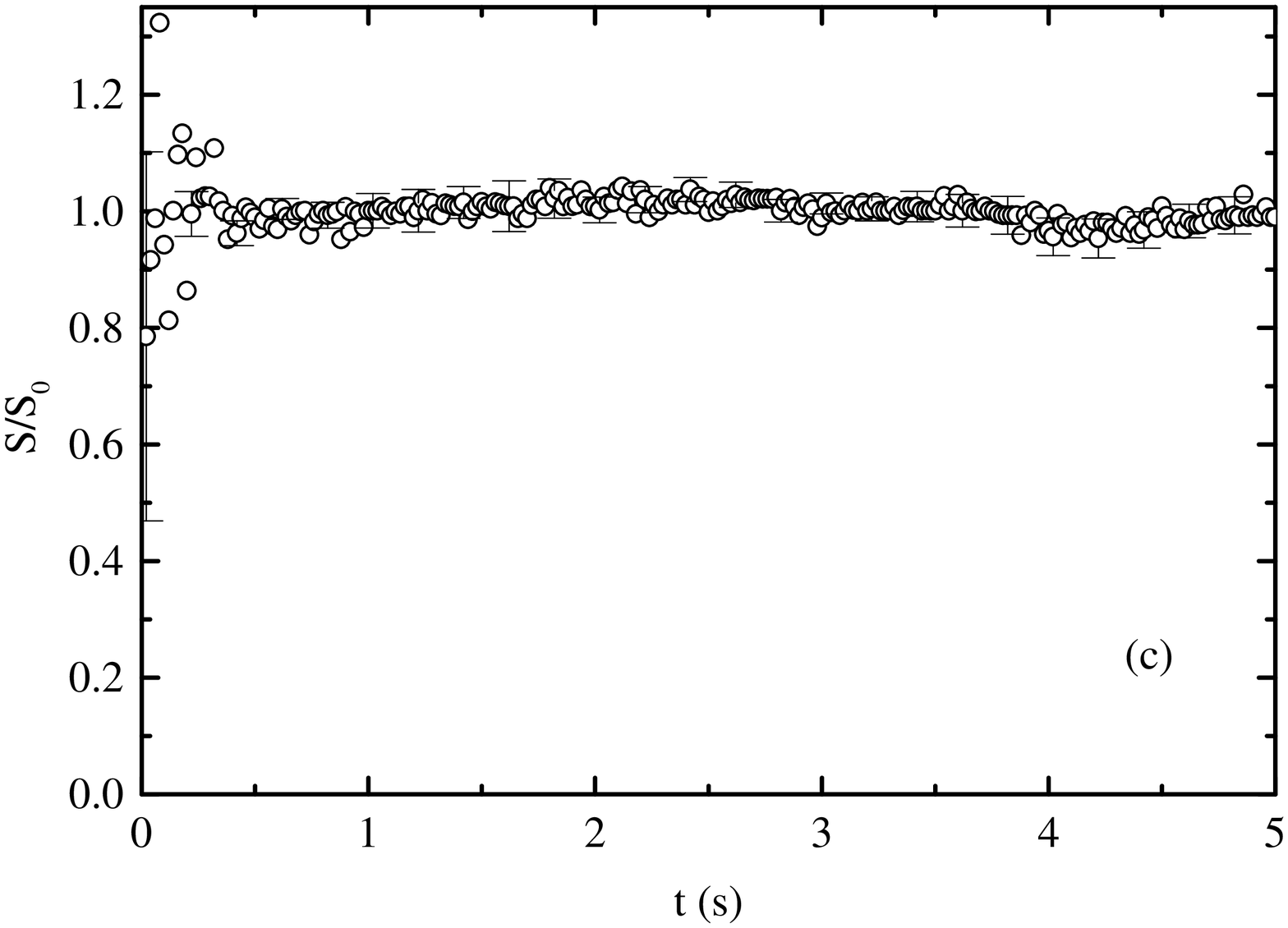}
\caption{Second-sound pulse amplitude $S/S_0$ as a function of time
following pressure quenches through the $\lambda$ transition:
(a) starting close to the transition, with
\trajectory{1.81}{30.3}{2.03}{6.2} ($\epsilon_i = -0.032$);
(b) starting further above the transition, with
\trajectory{2.05}{22.7}{2.09}{5.9} ($\epsilon_i = -0.089$);
(c) starting far above the transition, with
\trajectory{1.96}{34.1}{2.07}{6.1} ($\epsilon_i = -0.167$). }
\label{Fig8}
\end{figure}

We were surprised, however, to discover that no detectable vorticity
was created even when the expansion trajectory passed through the
transition. Signal amplitudes measured just after two such expansions
are shown by the data points of Fig.~\ref{Fig8}. It is immediately
evident that, unlike the results obtained from the original cell
\cite{H}, there is now no evidence of any systematic growth of the
signal amplitude with time or, correspondingly, for the creation of any
vortices at the transition. One possible reason is that the density of
vortices created is smaller than the theoretical estimates
\cite{C,D,E}, but we must also consider the possibility that they are
decaying faster than they can be measured.

\section{THE DECAY OF A VORTEX TANGLE} 
\begin{figure}[h]
\vspace{8in}
\includegraphics{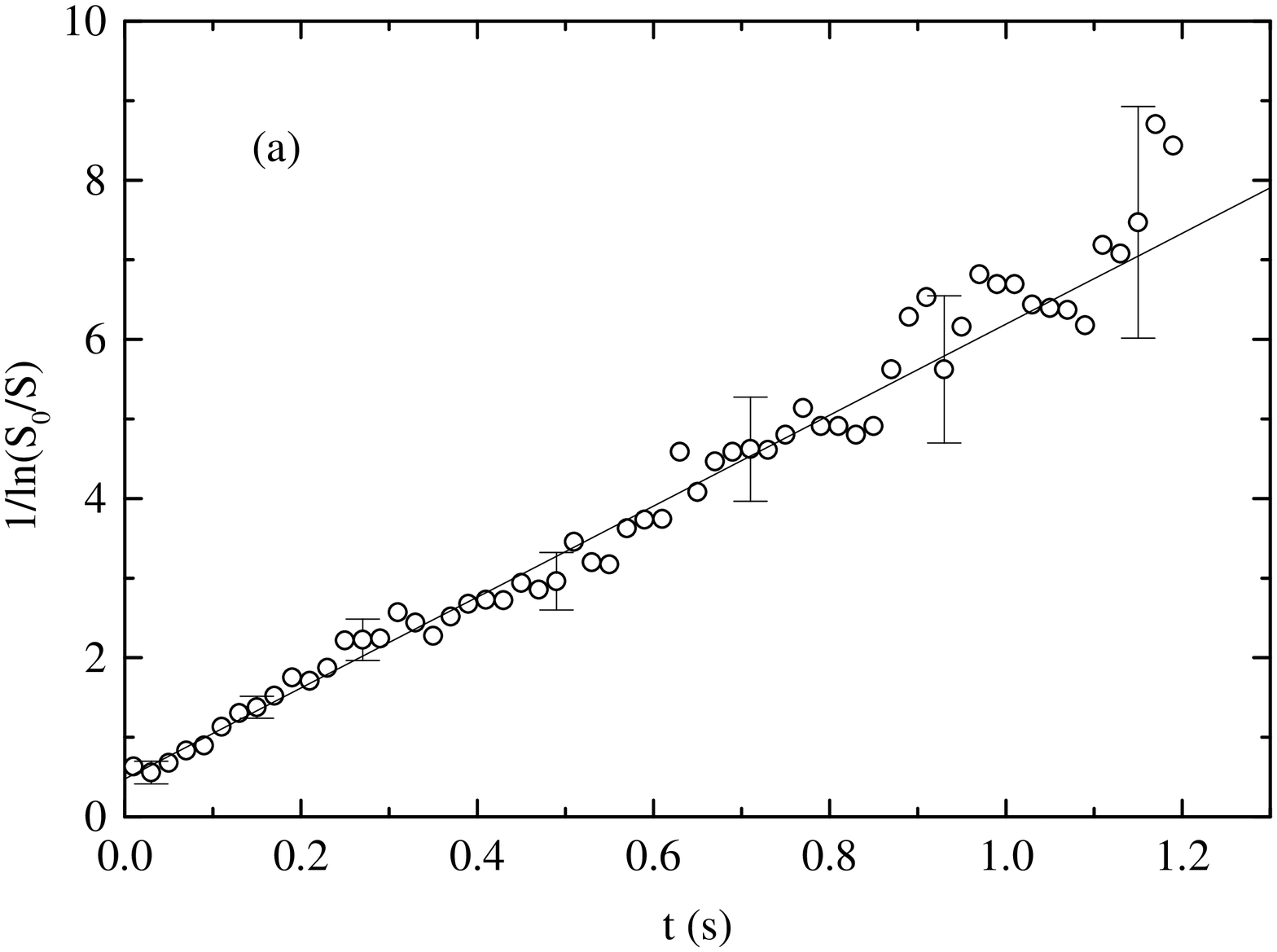}
\includegraphics{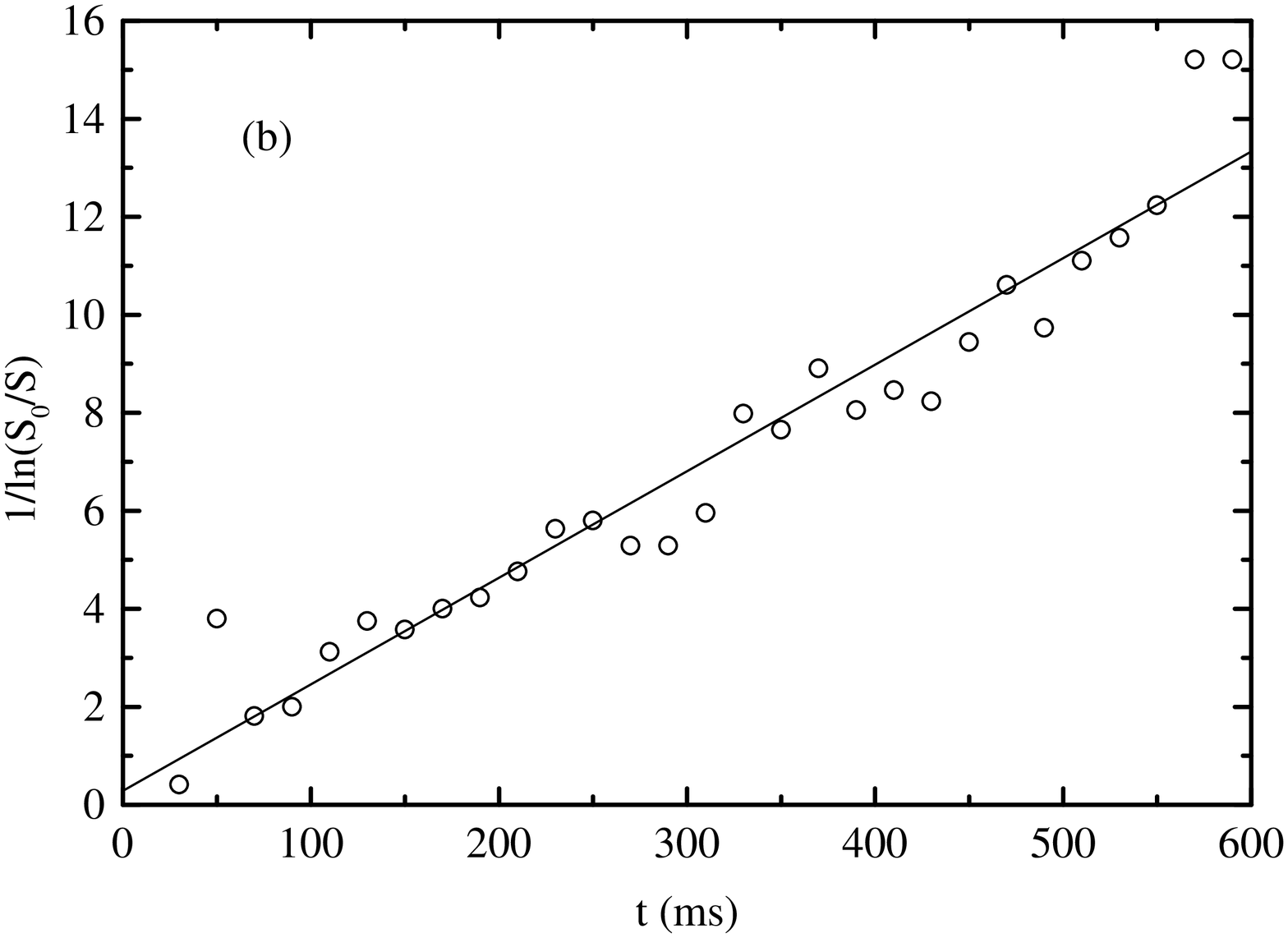}
\caption{Examples of $({\rm ln} S_0 - {\rm ln} S)^{-1}$ plotted
{\it versus} time $t$ for hydrodynamically created vortices.
}
\label{Fig9}
\end{figure}

\begin{figure}[h]
\vspace{7in}
\includegraphics{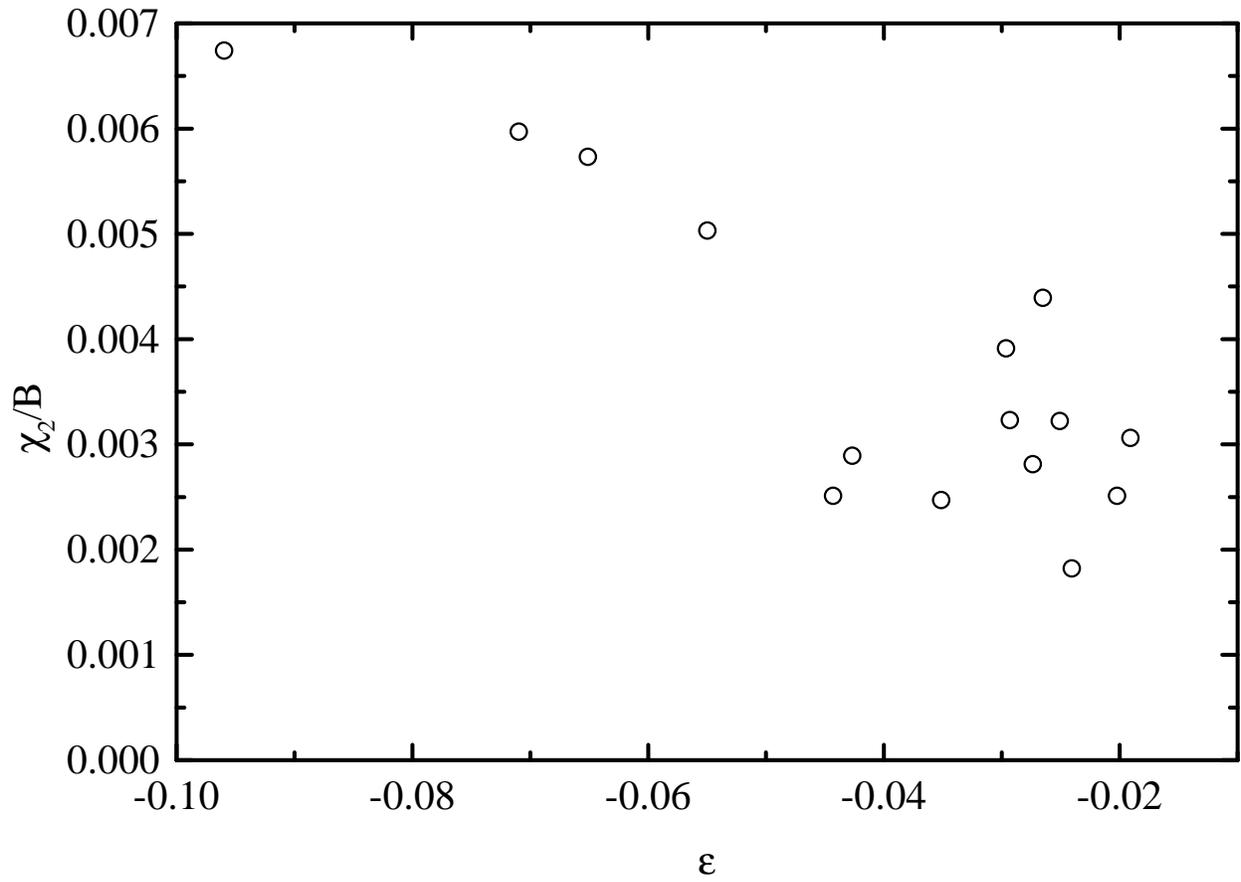}
\caption{Measured values of the parameter $\chi_2/B$ as a function of
distance $\epsilon$ from the $\lambda$-transition.}
\label{Fig10}
\end{figure}

To try to clarify matters, we
performed a subsidiary set of experiments, deliberately creating
vortices by conventional means and then following their decay by
measurements of the recovery of the second-sound signal amplitude. By
leaving the needle-valve open, so that $\sim 0.2\,{\rm cm}^3$ of liquid
from the dead volume outside the needle-valve actuator-bellows squirted
into the cell during an expansion, we could create large densities of
vorticity and observe their decay. Despite the highly non-equilibrium
situation that arises immediately following the expansion, as liquid
squirts into the cell, it was found that the temperature reached a
steady value after $\sim 6\,\rm ms$. Fig.~\ref{Fig9} shows two examples
of attenuation plots ($[\ln(S/S_0)]^{-1}$ against time $t$) and
these always, within the experimental errors, had the linear form
predicted by equation~\ref{eqnLine}. Fig.~\ref{Fig10} summarises the
results of a number of expansions from which it was possible to
determine $\chi_2/B$ as a function of temperature and pressure. We
found that $\chi_2/B$ was weakly temperature-dependent and over the
range of interest, $0.02<\epsilon<0.06$ it could be approximated by
$\chi_2/B=0.004\pm0.001$.

\begin{figure}[h]
\vspace{7in}
\includegraphics{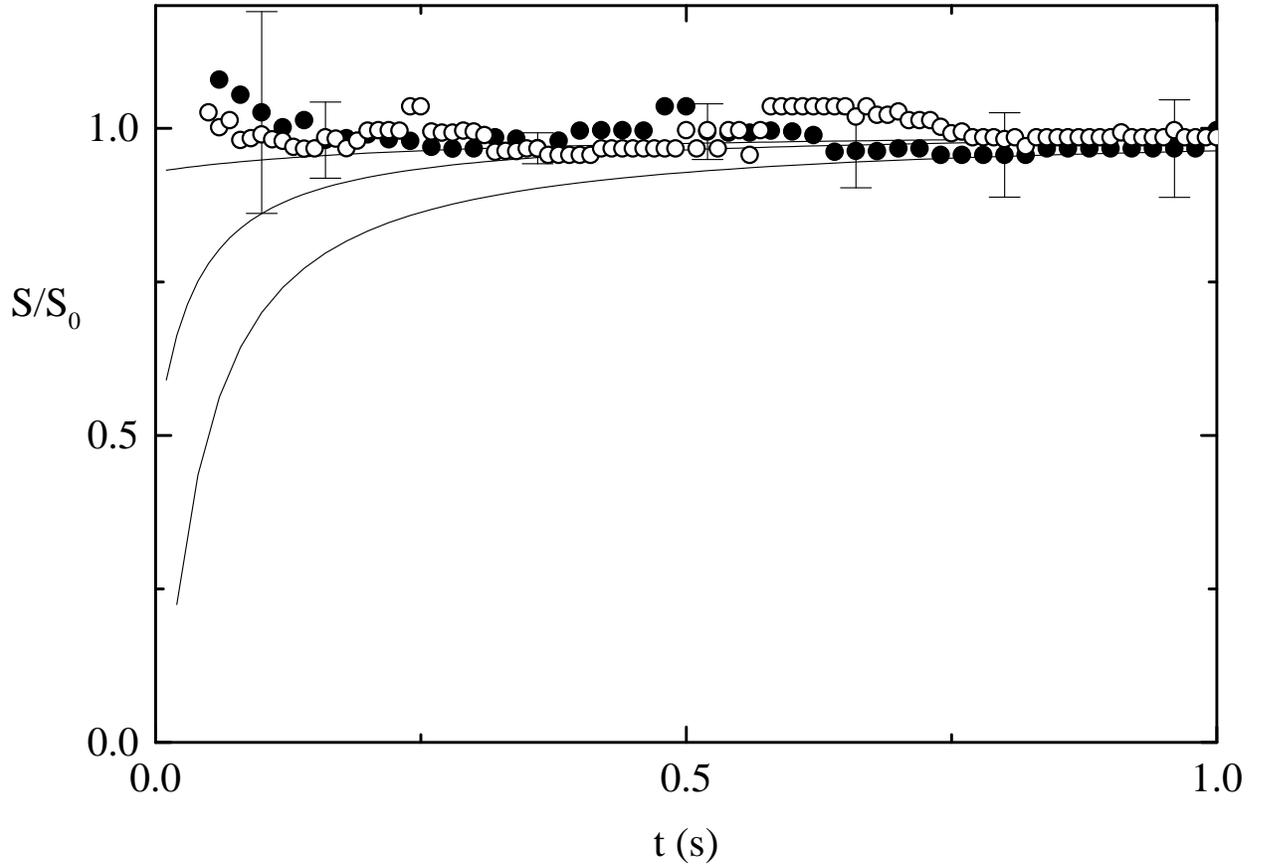}
\caption{Evolution of the second-sound amplitude $S$
with time, following an expansion of the cell at $t=0$ (data points),
normalised by its vortex-free value $S_0$. The open and closed symbols
correspond to signal repetition rates of 100 and 50 Hz respectively,
and in each case the starting and finishing conditions were $\epsilon_i
= -0.032, \epsilon_f = 0.039$. The curves refer to calculated signal
evolutions for different initial line densities, from the bottom, of
$10^{12}, 10^{11}$ and $10^{10}$ m$^{-2}$.}
\label{Fig11}
\end{figure}

\noindent This measured value of $\chi_2/B$ was then used to calculate
the evolution of $S/S_0$ with time for different values of $L_1$,
assuming $B=1$, yielding the curves shown in Fig.~8. From the
$\tau_{\rm q}$ derived from the gradient in Fig~\ref{Fig6}, and
Zurek's estimate (based on renormalisation-group theory) of
\begin{equation}
L_{\rm RG}=\frac{L_0}{\left(\tau_{\rm q}/\tau_0\right)^{2/3}}
\quad{\rm where}\quad
L_0=1.2\times10^{12}\,{\rm m}^{-2},\tau_0=100\,\rm ms
\end{equation}
\noindent we are thus led to expect that
$L_1\approx4\times10^{12}\,{\rm m}^2$. Direct comparison of the
calculated curves and measured data in Fig.~\ref{Fig11} shows that
this is plainly not the case. In fact, the data suggest that $L_1$, the
vorticity created by the transition, is no more than $10^{10}\,\rm
m^{-2}$, smaller than the expected value by at least two
orders-of-magnitude.

\section{DISCUSSION}

Given the apparently positive outcome of the earlier
investigations\cite{H}, the null result of the present experiment has
come as something of a surprise. There are several points to be made.
First, Zurek did not expect his estimates of $L_1$ be accurate to
better than one, or possibly two, orders-of-magnitude, and his more
recent estimate \cite{P} suggests somewhat lower defect-densities. So
it remains possible that his picture \cite{C,D,E,P} is correct for
$^4$He in all essential details, and that an improved experiment with
faster expansions now being planned will reveal evidence of the
Kibble-Zurek mechanism at work in liquid $^4$He. Secondly, it must be
borne in mind that (\ref{eqnVinen}), and the value of $\chi/B$ measured
(from plots like those in Fig.\ \ref{Fig10}) from the data of Fig.\
\ref{Fig9}, refer to hydrodynamically generated vortex lines. Vorticity
generated in the nonequilibrium phase transition might perhaps be
significantly different, e.g.\ in respect of its loop-size distribution
\cite{gaw}. It could therefore decay faster, and might consequently be
unobservable in the present experiments. Thirdly, it is surprising to
us that the $^3$He experiments \cite{I,J} seem to agree with Zurek's
original estimates \cite{C,D,E} whereas the present experiment shows
that they overstate $L_1$ by {\it at least} two orders-of-magnitude. It
is not yet known for sure why this should be, although an interesting
explanation of the apparent discrepancy has recently been suggested
\cite{Q} by Karra and Rivers. They suggest that fluctuations near to
$T_{\lambda}$ may change the winding number, i.e.\ reduce the density
of vortices produced in the $^4$He experiment, and they show that the
analogous density reduction would be much smaller in the case of the
$^3$He experiments.

\section*{ACKNOWLEDGMENTS} We acknowledge valuable discussions or
correspondence with S.N. Fisher, A.J. Gill, R.A.M.\ Lee, R.J. Rivers,
W.F. Vinen G.A.\ Williams and W.H. Zurek. The work was supported by the
Engineering and Physical Sciences Research Council (U.K.), the European
Commission and the European Science Foundation.

\newpage

%
%

\newpage

%
%

\newpage

%
%

\newpage

%
%

\newpage

%
%

\newpage

%
%

\newpage

%
%

\newpage

%
%

\newpage

%
%

\newpage

%
%

\newpage

%
%

\end{document}